\documentclass[3p,twocolumn]{elsarticle}

\usepackage{hyperref}
\usepackage{enumerate} 
\usepackage{amssymb}
\usepackage{algorithmicx,algorithm}
\usepackage{hyperref}
\usepackage{multirow}
\usepackage{stfloats}
\usepackage{graphicx}
\usepackage{color}
\usepackage[justification=centering]{caption}
\usepackage{url}
\usepackage{xcolor}
\usepackage{pdflscape}
\journal{Journal of \LaTeX\ Templates}
 \usepackage{pdfpages}
 \usepackage{graphicx}

\begin{document}

\begin{frontmatter}

\title{EBHI-Seg: A Novel Enteroscope Biopsy Histopathological Haematoxylin and Eosin Image Dataset for Image Segmentation Tasks}

\author[1]{Liyu Shi}

\author[2]{Xiaoyan Li\corref{cor1}}
\ead{lixiaoyan@cancerhosp-ln-cmu.com}

\author[1]{Weiming Hu}

\author[1]{Haoyuan Chen}

\author[1]{Jing Chen}

\author[1]{Zizhen Fan}

\author[1]{Minghe Gao}

\author[1]{Yujie Jing}

\author[1]{Guotao Lu}

\author[1]{Deguo Ma}

\author[1]{Zhiyu Ma}

\author[1]{Qingtao Meng}

\author[1]{Dechao Tang}

\author[3]{Hongzan Sun}

\author[4,5]{Marcin Grzegorzek }

\author[1]{Shouliang Qi}

\author[1]{Yueyang Teng}

\author[1]{Chen Li\corref{cor1}}

\cortext[cor1]{Corresponding author:} \ead{lichen@bmie.neu.edu.cn}

\address[1]{Microscopic Image and Medical Image Analysis Group, 
College of Medicine and Biological Information Engineering, 
Northeastern University, Shenyang 110169, China}

\address[2]{Department of Pathology, Cancer Hospital of China Medical University, 
Liaoning Cancer Hospital and Institute, Shengyang 110042, China}

\address[3]{Shengjing Hospital, China Medical University, Shenyang 110122, China}

\address[4]{Institute of Medical Informatics, University of Luebeck, Luebeck 23538, Germany}

\address[5]{Department of Knowledge Engineering, University of Economics in Katowice, 
Bogucicka 3, 40287 Katowice, Poland}

\begin{abstract}
\textbf{Background and Purpose:} Colorectal cancer is a  common fatal malignancy, the fourth 
most common cancer in men, and the third most common cancer in women worldwide. 
Timely detection of cancer in its early stages is essential for treating the disease. 
Currently, there is a lack of datasets for histopathological image segmentation of 
rectal cancer, which often hampers the assessment accuracy when computer technology 
is used to aid in diagnosis.

\textbf{Methods:} This present study provided a new publicly available 
\emph{Enteroscope Biopsy Histopathological Hematoxylin and Eosin  Image Dataset for 
Image Segmentation Tasks} (EBHI-Seg). To demonstrate the validity and extensiveness 
of EBHI-Seg, the experimental results for EBHI-Seg are evaluated using classical 
machine learning methods and deep learning methods.

\textbf{Results:} The experimental results showed that  deep learning methods had a better 
image segmentation performance when utilizing EBHI-Seg. The maximum accuracy of the 
Dice evaluation metric for the classical machine learning method is 0.948, while the 
Dice evaluation metric for the deep learning method is 0.965.

\textbf{Conclusion:} This publicly available dataset contained 5,170 images of six types of 
tumor differentiation stages and the corresponding ground truth images. The dataset 
can provide researchers with new segmentation algorithms for medical diagnosis of 
colorectal cancer, which can be used in the clinical setting to help doctors and patients.  
EBHI-Seg is publicly available at: https://doi.org/10.6084/m9.figshare.21540159.v1
\end{abstract}


\end{frontmatter}

\section{Introduction}

Colon cancer is a common deadly malignant tumor, the fourth most common cancer in men, 
and the third most common cancer in women worldwide. Colon cancer is responsible for 
$10\%$ of all cancer cases~\cite{sung2021global}. According to  prior research, colon 
and rectal tumors share many of the same or similar characteristics. Hence, they are 
often classified collectively. The present study categorized rectal and colon cancers 
into one colorectal cancer category~\cite{pamudurthy2020advances}. Histopathological 
examination of the intestinal tract is both the gold standard for the diagnosis of 
colorectal cancer and a prerequisite for disease treatment~\cite{thijs1996diagnostic}.

The advantage of using the intestinal biopsy method to remove a part of the intestinal 
tissue for  histopathological analysis, which is used to determine the true status of 
the patient, is that it considerably reduces damage to the body and rapid wound 
healing~\cite{labianca2013early}. The histopathology sample is then sectioned and 
processed with Hematoxylin and Eosin (H\&E). Treatment with H\&E is a common approach 
when staining tissue sections to show the inclusions between the nucleus and cytoplasm 
and highlight the fine structures between 
tissues~\cite{fischer2008hematoxylin,chan2014wonderful}. When a pathologist performs 
an examination of the colon, they first examine the histopathological sections for 
eligibility and find the location of the lesion. The pathology sections are then examined 
and diagnosed using a low magnification microscope. If finer structures need to be 
observed, the microscope is adjusted to use high magnification for further analysis. 
However, the following problems usually exist in the diagnostic process: the diagnostic 
results become more subjective and varied due to different doctors reasons; doctors can 
easily overlook some information in the presence of a large amount of test data; it is 
difficult to analyze large amounts of previously collected data~\cite{gupta2021breast}. 
Therefore, it is a necessary to address these issues effectively.

With the development and popularization of computer-aided diagnosis (CAD), the 
pathological sections of each case can be accurately and efficiently examined with 
the help of computers~\cite{mathew2021computational}. Now, CAD is widely used in 
many biomedical image analysis tasks, such as 
microorganism image analysis~\cite{li2019survey,zhang2022applications,zhang2021lcu,zhao2022comparative,
kulwa2022new,ma2022state,kulwa2023segmentation,zhang2022comprehensive,zhang2021comprehensive}, 
COVID-19 image analysis~\cite{rahaman2020identification}, 
histopatholgical image analysis~\cite{chen2022gashis,li2022hierarchical,hu2022gashissdb,
chen2022mcam,hu2022comparative,li2022comprehensive,sun2020gastric,li2022comprehensives}, 
cytopathological image analysis~\cite{rahaman2020survey,mamunur2021deepcervix,liu2021aspect,liu2022cvm} 
and sperm video analysis~\cite{chen2022svia,zou2022tod}. 
Therefore, the application of  computer vision technology for colorectal cancer CAD provides 
a new direction in this research field~\cite{pacal2020comprehensive}.

One of the fundamental features of CAD is the aspect of image segmentation, the results 
of which can be used as key evidence in the pathologists' diagnostic processes. Along 
with the rapid development of medical image segmentation methodology, there is a wide 
demand for its application to identify benign and malignant tumors, tumor differentiation 
stages, and other related fields~\cite{miranda2016survey}. Therefore, a multi-class 
image segmentation method is needed to obtain high segmentation accuracy and good 
robustness~\cite{kotadiya2019review}.

The present study presents a novel \emph{Enteroscope Biopsy Histopathological H}\&\emph{E 
Image Dataset for Image Segmentation Tasks} (EBHI-Seg), which  contains 5,710 electron 
microscopic images of histopathological colorectal cancer sections that encompass six 
tumor differentiation stages: normal, polyp, low-grade intraepithelial neoplasia, 
high-grade intraepithelial neoplasia, serrated adenoma, and adenocarcinoma. The 
segmentation coefficients and evaluation metrics are obtained by segmenting the images 
of this dataset using different classical machine learning methods and novel deep 
learning methods.

\section{Related Work}

The present study analyzed and compared the existing colorectal cancer biopsy dataset 
and provided an in-depth exploration of the currently known research findings. The 
limitations of the presently available colorectal cancer dataset were also pointed out.

The following conclusions were obtained in the course of the study. For existing datasets, 
the data types can be grouped into two major categories: Multi and Dual Categorization 
datasets. Multi Categorization datasets contain tissue types at all stages from Normal 
to Neoplastic. In~\cite{trivizakis2021neural}, a dataset called ``Collection of textures 
in colorectal cancer histology'' is described. It includes 5,000 patches of size 
74 $\mu$m $\times $ 74 $\mu$m and contains seven categories. However, because there were 
only 10 images, it is too small for a data sample and lacked generalization capability. 
In \cite{chen2022mcam}, a dataset called ``NCT-CRC-HE-100K'' is proposed. This is a set 
of 100,000 non-overlapping image patches of histological human colorectal cancer (CRC) 
and normal tissue samples stained with (H\&E) that was presented by the National Center 
for Tumor Diseases (NCT). These image patches are from nine different tissues with an 
image size of 224 $\times $ 224 pixels. The nine tissue categories are adipose, background, 
debris, lymphocytes, mucus, smooth muscle, normal colon mucosa, cancer-associated stroma, 
and colorectal adenocarcinoma epithelium. This dataset is publicly available and commonly 
used. However, because the image sizes are all 224 $\times $ 224 pixels, the dataset 
underperformed in some global details that need to be observed in individual categories. 
Two datasets are utilized in \cite{oliveira2021cad}: one containing colonic 
H\&E-stained biopsy sections (CRC dataset) and the other consisting of prostate cancer 
H\&E-stained biopsy sections (PCa dataset). The CRC dataset contains 1,133 colorectal 
biopsy and polypectomy slides grouped into three categories and labelled as non-neoplastic, 
low-grade and high-grade lesions. In \cite{kausar2021sa}, a dataset named ``MICCAI 2016 
gland segmentation challenge dataset (GlaS)'' is used. This dataset contained 165 
microscopic images of H\&E-stained colon glandular tissue samples, including 85 training 
and 80 test datasets. Each dataset is grouped into two parts: benign and malignant tumors. 
The image size is 775 $\times $ 522 pixels. Since this dataset has only two types of data 
and the number of data is too little, so that it performs poorly on some multi-type training.

Dual Categorization datasets usually contain only two types of tissue types: Normal and 
Neoplastic. In \cite{wei2021learn}, a dataset named ``FFPE'' is proposed. This dataset 
obtained its images by extracting 328 Formalin-fixed Paraffin-embedded (FFPE) whole-slide 
images of colorectal polyps classified into two categories of : hyperplastic polyps (HPs) 
and sessile serrated adenomas (SSAs). This dataset contained 3,125 images with an image 
size of 224 $\times $ 224 pixels and is small in type and number. In~\cite{bilal2022ai}, 
two datasets named ``UHCW'' and ``TCGA'' are proposed. The first dataset is a colorectal 
cancer biopsy sequence developed at the University Hospital of Coventry and Warwickshire 
(UHCW) for internal validation of the rectal biopsy trial. The second dataset is the 
Cancer Genome Atlas (TCGA) for external validation of the trial. This dataset is commonly 
used as a publicly available cancer dataset and stores genomic data for more than 20 types 
of cancers. The two dataset types are grouped into two categories: Normal and Neoplastic. 
The first dataset contains 4,292 slices, and the second dataset contained 731 slices with 
an image size of 224 $\times $ 224 pixels.

All of the information for the existing datasets is summarized in Table~\ref{Table:compare3}. 
The issues associated with the dataset mentioned above included fewer data types, small 
amount of data, inaccurate dataset ground truth, etc. The current study required an 
open-source multi-type colonoscopy biopsy image dataset.
    
\begin{table*}[htbp!]
\small
\centering
\caption{  A dataset for the pathological classification of colorectal cancer. }
\label{Table:compare3}
\resizebox{\linewidth}{!}{
\begin{tabular}{ccccccccc}
    \hline
         
                  &   \textbf{ Dataset Name }  & \textbf{ Multi Categorization } & \textbf{ Amount}  & \textbf{ Size} & \textbf{ Year }\\
\hline
                  &{\multirow{-4}*{\shortstack{Collection of textures in\\colorectal cancer histology}}}   & {\shortstack{lymphoid follicles, mucosal glands,\\debris, adipose, tumor epithelium\\simple stroma, complex stroma,\\background patches with no tissue}} & {\multirow{-4}*{5000}}              & 
                 {\multirow{-4}*{\shortstack{74$\mu$m $\times $ 74$\mu$m\\(0.495 micrometre\\per pixel)}}}      & {\multirow{-4}*{2016}}   \\ 

\hline
    &{\multirow{-2}*{\shortstack{HE-NCT-CRC-100K}}}   & {\shortstack{MUS, NORM, STR, TUM\\ADI, BACK, DEB, LYM, MUC}} & {\multirow{-2}*{100000}}    &  {\multirow{-2}*{224$\times $224 pixels}}    & {\multirow{-2}*{2016}}       \\

\hline
   & {\shortstack{MICCAI’16 gland seg-\\mentation challenge dataset}}   & {\multirow{-2}*{\shortstack{benign tumors, malignant tumors}}} & {\multirow{-2}*{85}}  &  {\multirow{-2}*{775$\times $522 pixels}}    & {\multirow{-2}*{2017}}       \\

      \hline
   & {\multirow{-2}*{\shortstack{CRC dataset}}}   & {\shortstack{non-neoplastic, low-grade,\\high-grade lesions	}} & {\multirow{-2}*{1133}}  &  {\multirow{-2}*{512$\times $512 pixels}}     & {\multirow{-2}*{2021}}       \\
    \hline
       \hline
         
                  &   \textbf{ Dataset Name }  & \textbf{ Dual Categorization } & \textbf{ Amount}  & \textbf{ Size} & \textbf{ Year }\\
\hline
   & {\shortstack{FFPE}}   & {\shortstack{HPs, SSAs	}} & 3152 &  224$\times $224 pixels    & 2021       \\

\hline
    &{\shortstack{The Cancer Genome\\Atlas dataset}}   & {\multirow{-1.5}*{\shortstack{Normal, Neoplastic	}}} & {\multirow{-1.5}*{731}}  &  {\multirow{-1.5}*{224$\times $224 pixels }}    & {\multirow{-1.5}*{2021}}      \\
\hline
   & {\shortstack{University Hospitals\\Coventry and Warwick-\\shire dataset}}   & {\multirow{-3}*{\shortstack{Normal, Neoplastic	}}} & {\multirow{-3}*{4292}}  &  {\multirow{-3}*{224$\times $224 pixels}}     & {\multirow{-3}*{2021}}      \\
   
\hline     
\end{tabular}
}
\end{table*}

\section{Basic Information for EBHI-Seg}
\label{section:basic}
\subsection{Dataset Overview}
The dataset in the present study contained 5,710 histopathology images, including 
2,855 histopathology section images and 2,855 ground truth images. The basic information 
for the dataset is described in detail below. EBHI-Seg is publicly available at: 
https://doi.org/10.6084/m9.figshare.21540159.v1

In the present paper, H\&E-treated histopathological sections of colon tissues are used 
as data for evaluating image segmentation. The dataset is obtained from two 
histopathologists at the Cancer Hospital of China Medical University (proved by 
``Research Project Ethics Certification'' (No. 202229) ). It is prepared by 12 
biomedical researchers according to the following rules: Firstly, if there is only 
one differentiation stage in the image and the rest of the image is intact, then the 
differentiation stage became the image label; Secondly, if there is more than one 
differentiation stage in the image, then the most obvious differentiation is selected 
as the image label; In general, the most severe and prominent differentiation in the 
image was used as the image label.

Intestinal biopsy was used as the sampling method in this dataset. The magnification 
of the data slices is $400\times $, with an eyepiece magnification of $10\times $ and 
an objective magnification of $40\times $. A Nissan Olympus microscope and NewUsbCamera 
acquisition software are used. The image input size is $224\times 224$ pixels, and the 
format is *.png. The data are grouped into five types described in detail in section 2.2.

\subsection{Data Type Description}

\begin{figure*}[htbp!]
\centering
\includegraphics[width= 0.95 \textwidth]{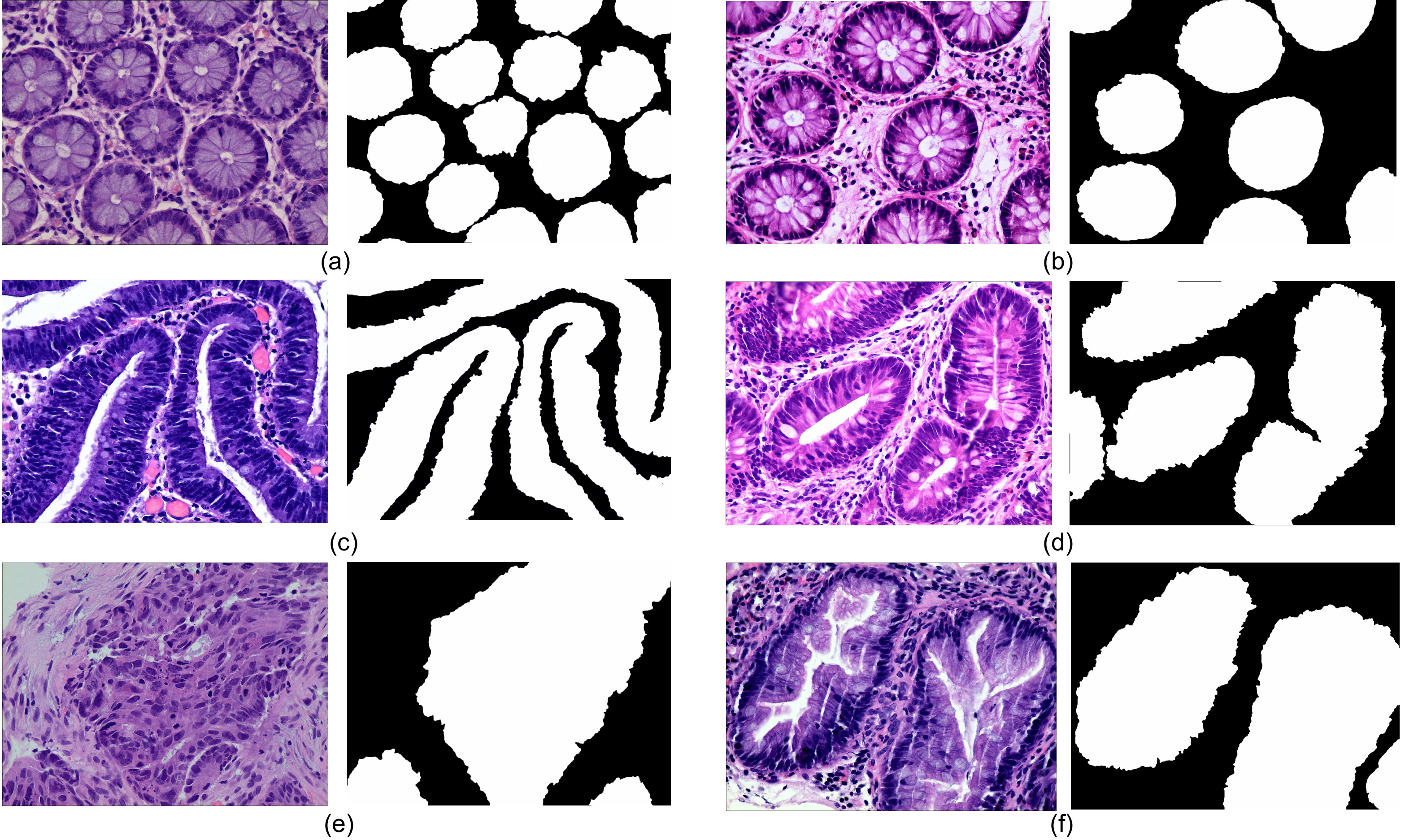}
\caption{An example of histopathological images database: 
(a) Normal and ground truth, 
(b) Polyp and ground truth, 
(c) High-grade Intraepithelial Neoplasia and ground truth, 
(d) Low-grade Intraepithelial Neoplasia and ground truth, 
(e) Adenocarcinoma and ground truth, 
(f) Serrated adenoma and ground truth.}     
\label{fig:Normal}
\end{figure*}

Normal: Colorectal tissue sections of the standard category are made-up of consistently 
ordered tubular structures and that does not appear infected  when viewed under a light 
microscope~\cite{de2001pathology}. Section images with the corresponding ground truth 
images are shown in Figure~\ref{fig:Normal}(a).

Polyp: Colorectal polyps are similar in shape to the structures in the normal category, 
but have a completely different histological structure. A polyp is a redundant mass that 
grows on the surface of the body's cells. Modern medicine usually refers to polyps as 
unwanted growths on the mucosal surface of the body~\cite{cooper1998pathology}. The 
pathological section of the polyp category also has an intact luminal structure with 
essentially no nuclear division of the cells. Only the atomic mass is slightly higher 
than that in the normal category. The polyp category and corresponding ground truth images 
are shown in Figure~\ref{fig:Normal}(b).

Intraepithelial neoplasia: Intraepithelial neoplasia (IN) is the most critical 
precancerous lesion. Compared to the normal category, its histological images show 
increased branching of adenoid structures, dense arrangement, and different luminal 
sizes and shapes. In terms of cellular morphology, the nuclei are enlarged and vary in 
size, while nuclear division increases ~\cite{ren2013missed}. The standard Padova 
classification currently classifies intraepithelial neoplasia into low-grade and 
high-grade INs. High-grade IN demonstrate more pronounced structural changes in the 
lumen and nuclear enlargement compared to low-grade IN. The images and ground truth 
diagrams of high-grade and low-grade INs are shown in Figure~\ref{fig:Normal}(c)(d).

Adenocarcinoma: Adenocarcinoma is a malignant digestive tract tumor with a very irregular 
distribution of luminal structures. It is difficult to identify its border structures 
during observation, and the nuclei are significantly enlarged at this 
stage~\cite{jass2012histological}. An adenocarcinoma with its corresponding ground 
truth diagram is shown in Figure~\ref{fig:Normal}(e).

Serrated adenoma: Serrated adenomas are uncommon lesions, accounting for 1\% of all 
colonic polyps ~\cite{spring2006high}. The endoscopic surface appearance of serrated 
adenomas is not well characterized but is thought to be similar to that of colonic 
adenomas with tubular or cerebral crypt openings~\cite{li2007histopathology}. The image 
of a serrated adenoma with a corresponding ground truth diagram is shown in 
Figure~\ref{fig:Normal}(f).

\section{Evaluation of EBHI-Seg}
\label{section:evaluation}
\subsection{Image Segmentation Evaluation Metric}

Six evaluation metrics are commonly used for image segmentation tasks. The Dice ratio 
metric is a standard metric used in medical images that is often utilized to evaluate 
the performance of image segmentation algorithms. It is a validation method based on 
spatial overlap statistics that measures the similarities between the algorithm 
segmentation output and ground truth~\cite{zou2004statistical}. The Dice ratio is 
defined in Eq. (1).
\begin{equation}
\centering
\rm DiceRatio = \frac{2\left | X\cap  Y\right | }{\left | X \right | +  \left | Y \right |  }.
\end{equation}

In Eq. (1), for a segmentation task, $X$ and $Y$ denote the ground truth and segmentation 
mask prediction, respectively. The range of the calculated results is [0,1], and the 
larger the result the better.

The Jaccard index is a classical set similarity measure with many practical applications 
in image segmentation. The Jaccard index measures the similarity of a finite set of 
samples: the ratio between the intersection and concatenation of the segmentation results 
and ground truth~\cite{jaccard1901etude}. The Jaccard index is defined in Eq. (2).
\begin{equation}
\rm JaccardIndex = \frac{\left | X\cap Y \right | }{\left | X\cup Y \right | }.
\end{equation}
The range of the calculated results is [0,1], and the larger the result the better.

Recall and precision are the recall and precision rates, respectively. The range of 
the calculated results is [0,1]. A higher output indicates a better segmentation result. 
Recall and precision are defined in Eq. (3) and Eq. (4),
\begin{equation}
\rm Precison = \frac{TP}{TP+FP} ,
\end{equation}

\begin{equation}
\rm Recall = \frac{TP}{TP+FN} ,
\end{equation}
where TP, FP, TN, and FN are defined in table~\ref{Table:compare2}.

\begin{table}[ht] \small
\caption{Confusion Matrix}
\centering
\label{Table:compare2}
\begin{tabular}{c|c|c}
\hline
\multirow{2}{*}{Ground truth} & \multicolumn{2}{c}{Predict mask}    \\
\cline{2-3}
                              & Positive  & Negative \\ 
                              \hline
Positive              & TP                & TN              \\
\hline
Negative               & FP                & FN     \\
\hline        
\end{tabular}
\end{table}

The conformity coefficient (Confm Index) is a consistency coefficient, which is calculated 
by putting the binary classification result of each pixel from [$-\infty$,1] into continuous 
interval [$-\infty$,1] to calculate the ratio of the number of incorrectly segmented pixels 
to the number of correctly segmented pixels to measure the consistency between the 
segmentation result and ground truth. The conformity coefficient is defined in Eq. (5)(6),
\begin{equation}
\rm ConfmIndex = (1-\frac{\theta _{AE}  }{\theta _{TP}} ) , \theta _{TP}>0 ,
\end{equation}
\begin{equation}
\rm ConfmIndex = Failure , \theta _{TP}=0 ,
\end{equation}

where $\theta _{AE}$= $\theta _{FP}$+$\theta _{FN}$ represents all errors of the fuzzy 
segmentation results. $\theta _{TP}$ is the number of correctly classified pixels. 
Mathematically, ConfmIndex can be negative infinity if $\theta _{TP}$=0. Such a 
segmentation result is definitely inadequate and treated as failure without the need 
of any further analysis.

\subsection{Classical Machine Learning Methods}    
\begin{figure*}[htbp!]
\centering
\includegraphics[width= 0.95 \textwidth]{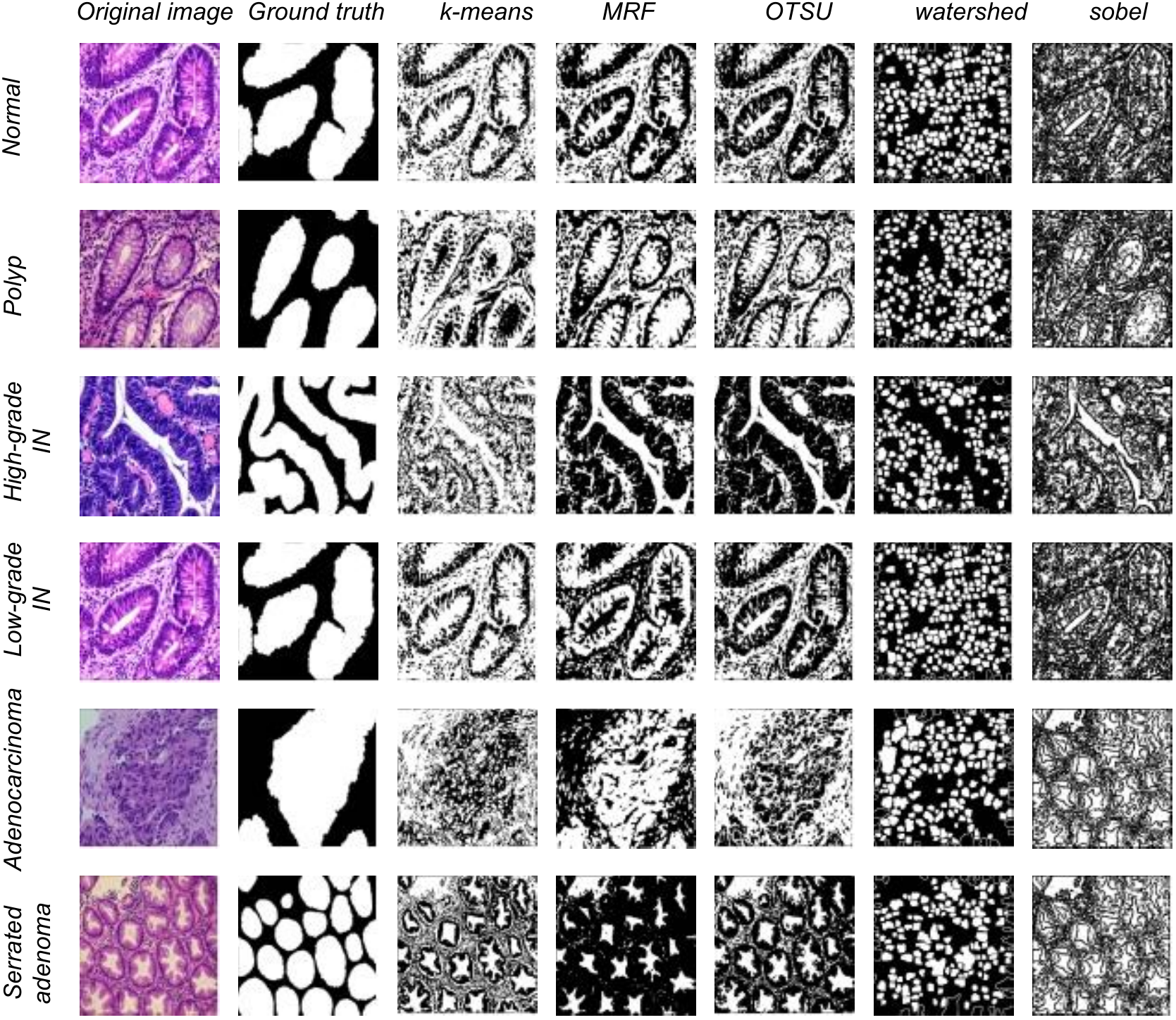}
\setlength{\abovecaptionskip}{1.0cm}
\caption{Five types of data segmentation results obtained by different classical machine learning methods.}     
\label{fig:1}
\end{figure*}

Image segmentation is one of the most commonly used methods for classifying image pixels 
in decision-oriented applications~\cite{naz2010image}. It groups an image into regions 
high in pixel similarity within each area and has a significant contrast between 
different regions~\cite{zaitoun2015survey}. Machine learning methods for segmentation 
distinguish the image classes using image features. 
(1) $k$-means algorithm is a classical division-based clustering algorithm, where image 
segmentation means segmenting the image into many disjointed regions. The essence is the 
clustering process of pixels, and the $k$-means method is one of the simplest clustering 
methods~\cite{dhanachandra2015image}. 
Image segmentation of the present study dataset is performed using the classical machine 
learning method described above. 
(2) Markov random field (MRF) is a powerful stochastic tool that models the joint 
probability distribution  of an image based on its local spatial 
action~\cite{deng2004unsupervised}. It can extract the texture features of the image and 
model the image segmentation problem. 
(3) OTSU algorithm is a global adaptive binarized threshold segmentation algorithm that 
uses the maximum inter-class variance between the image background and the target image 
as the selection criterion~\cite{huang2021otsu}. The image is grouped into foreground 
and background parts based on its grayscale characteristics independent of the brightness 
and contrast.
(4) Watershed algorithm is a region-based segmentation method, that takes the similarity 
between neighboring pixels as a reference and connects those pixels with similar spatial 
locations and grayscale values into a closed contour to achieve the segmentation 
effect~\cite{khiyal2009modified}. 
(5) Sobel algorithm has two operators, where one detects horizontal edges and the other 
detects vertical flat edges. An image is the final result of its operation. Sobel edge 
detection operator is a set of directional operators that can be used  to perform edge 
detection from different directions~\cite{zhang2012probe}. 
The segmentation results are shown in Figure~\ref{fig:1}.

The performance of EBHI-Seg for different machine learning methods is observed by 
comparing the images segmented using classical machine learning methods with the 
corresponding ground truth. The segmentation evaluation metrics results are shown in 
Table~\ref{Table:compare}. The Dice ratio algorithm is a similarity measure, usually 
used to compare the similarity of two samples. The value of one for this metric is c
onsidered to indicate the best effect, while the value of the worst impact is zero. 
The Table~\ref{Table:compare} shows that $k$-means has a good Dice ratio algorithm 
value of up to 0.650 in each category. The  MRF and Sobel segmentation results also 
achieved a good Dice ratio algorithm  value of around 0.6. In terms of image precision 
and recall segmentation coefficients, $k$-means is maintained at approximately 0.650 
in each category. In the classical machine learning methods, $k$-means has the best 
segmentation results, followed by MRF and Sobel. OTSU has a general effect, while 
the watershed algorithm has various coefficients that are much lower than those in 
the above methods. Moreover,there are apparent differences in the segmentation results 
when using the above methods.

\begin{table*}[htbp!] 
\small
\centering
\caption{Evaluation metrics for five different segmentation methods based on classical machine learning. }
\label{Table:compare}
\begin{tabular}{cccccccc}
\hline
                 &           & DiceRatio  & JaccardIndex  & ConformityCoefficient & Precision & Recall  \\
\hline
                 & $k$-means   & 0.648 & 0.488              & -0.184     & 0.646     & 0.663   \\
                 & MRF       & 0.636 & 0.473              & -0.230     & 0.637    & 0.658  \\
Normal           & OTSU      & 0.410 & 0.265              & -2.871     & 0.515     & 0.351   \\
                 & Watershed & 0.461 & 0.300              & -1.375     & 0.668     & 0.356   \\
                 & Sobel     & 0.652 & 0.487              & -0.102     & 0.763     & 0.579   \\
\hline
                 & $k$-means   & 0.592 & 0.430               & -0.528     & 0.546     & 0.663   \\
                 & MRF       & 0.511 & 0.362              & -2.133     & 0.540     & 0.502   \\
Polyp            & OTSU      & 0.400 & 0.259              & -3.108     & 0.413     & 0.399   \\
                 & Watershed & 0.433 & 0.277               & -1.675     & 0.551     & 0.362   \\
                 & Sobel     & 0.583 & 0.416              & -0.499     & 0.626     & 0.562   \\
\hline
                 & $k$-means  & 0.626 & 0.478              & -0.467     & 0.650     & 0.620   \\
                 & MRF       & 0.550 & 0.441              & -30.85     & 0.614     & 0.526   \\
High-grade IN    & OTSU      & 0.249 & 0.150              & -12.06     & 0.373     & 0.191   \\
                 & Watershed & 0.472 & 0.309              & -1.258     & 0.738     & 0.350   \\
                 & Sobel     & 0.634 & 0.469              & -0.200     & 0.728     & 0.577   \\
\hline
                 & $k$-means   & 0.650 & 0.492              & -0.172     & 0.651     & 0.663   \\
                 & MRF       & 0.554 & 0.404              & -1.808     & 0.643     & 0.504   \\
Low-grade IN     & OTSU     & 0.886 & 0.811              & 0.6998     & 0.832     & 0.979   \\
                 & Watershed & 0.464 & 0.303              & -1.345     & 0.676     & 0.357   \\
                 & Sobel     & 0.656 & 0.492              & -0.079     & 0.771     & 0.582   \\
\hline
                 & $k$-means   & 0.633 & 0.481               & -0.414     & 0.655     & 0.645   \\
                 & MRF       & 0.554 & 0.404               & -1.808     & 0.643     & 0.504   \\
Adenocarcinoma   & OTSU      & 0.336 & 0.215               & -5.211     & 0.454     & 0.282   \\
                 & Watershed & 0.458 & 0.298               & -1.437     & 0.700     & 0.349   \\
                 & Sobel     & 0.553 & 0.388              & -0.733     & 0.692     & 0.484   \\
\hline
                 & $k$-means   & 0.636 & 0.473               & -0.230     & 0.637    & 0.658  \\
                 & MRF       & 0.571 & 0.419               & -0.898     & 0.656     & 0.547   \\
Serrated adenoma & OTSU      & 0.393 & 0.248               & -2.444     & 0.565     & 0.315   \\
                 & Watershed & 0.449 & 0.290               & -1.494     & 0.656     & 0.345   \\
                 & Sobel     & 0.698 & 0.541              & 0.7484     & 0.662     & 0.572   \\  
\hline     
\end{tabular}
\end{table*}

In summary, EBHI-Seg has significantly different results when using different classical 
machine learning segmentation methods. Different classical machine learning methods 
have an obvious differentiation according to the image segmentation evaluation metrics. 
Therefore, EBHI-Seg can effectively evaluate the segmentation performance of different 
segmentation methods.

\subsection{Deep Learning Methods}
\begin{figure*}[htbp!]
\centering
\includegraphics[width= 0.95 \textwidth]{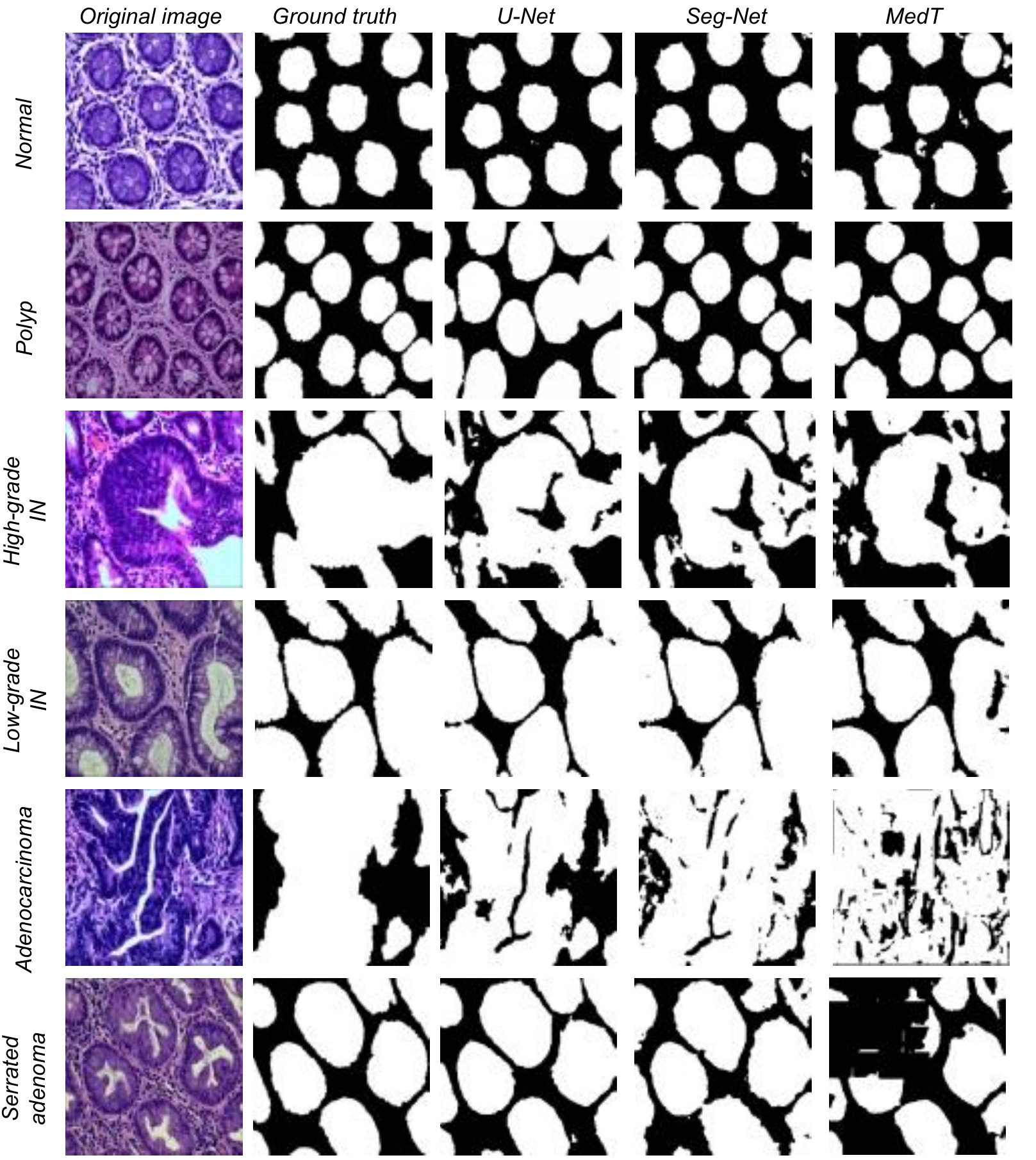}
\setlength{\abovecaptionskip}{1.5cm}
\caption{Three types of data segmentation results obtained by different deep learning methods.}     
\label{fig:2}
\end{figure*}

Besides the classical macine learning metheds tested above, some popular deep learning 
methods are also tested.
(1) Seg-Net is an open source project for image 
segmentation~\cite{badrinarayanan2017segnet}. The network is identical to the 
convolutional layer of VGG-16, with the removal of the fully-connected hierarchy and 
the addition of max-pooling indices resulting in improved boundary delineation. 
Seg-Net performs better in large datasets. 
(2) U-Net network structure was first proposed in 2015~\cite{ronneberger2015u} for 
medical imaging. U-Net is lightweight, and its simultaneous detection of local and 
global information is helpful for both information extraction and diagnostic results 
from clinical medical images. 
(3) MedT is a network published in 2021, which is a transformer structure that applies 
an attention mechanism based on medical image segmentation~\cite{valanarasu2021medical}. 
The segmentation results are shown in Figure~\ref{fig:2}.

The segmentation effect is test on the present dataset using three deep learning models. In the experiments, each model is trained using the ratio of the training set, validation set, and test set of $4:4:2$. The model learning rate is set to $3e-6$, epochs are set to 100, and batch-size is set to 1. The dataset segmentation results of using three different models are shown in Figure~\ref{fig:2}. The experimental segmentation evaluation metrics are shown in Table~\ref{Table:compare1}. Overall, deep learning performs much better than classical machine learning methods. Among them, the evaluation indexes of the training results using the U-Net and Seg-Net models can reach 0.90 on average. The evaluation results of the MedT model are slightly worse at a level, between 0.70 and 0.80. The training time is longer for MedT and similar for U-Net and Seg-Net.


\begin{table*}[htbp!] 
\small
\centering
\caption{Evaluation metrics for three different segmentation methods based on deep learning.  }
\label{Table:compare1}
\begin{tabular}{cccccccc}
\hline
                 &           & DiceRatio  & JaccardIndex  & ConformityCoefficient & Precision & Recall  \\
\hline
                 & U-Net   & 0.411 & 0.263               & -2.199     & 0.586     & 0.328   \\
Normal           & Seg-Net & 0.777 & 0.684               & -0.607    & 0.895    & 0.758  \\
                 & MedT    & 0.676 & 0.562               & -0.615     & 0.874     & 0.610   \\
\hline
                 & U-Net   & 0.965 & 0.308              & -1.514     & 0.496     & 0.470   \\
Polyp            & Seg-Net & 0.937 & 0.886              & 0.858     & 0.916    & 0.965  \\
                 & MedT    & 0.771 & 0.643              & 0.336     & 0.687     & 0.920   \\
\hline
                 & U-Net   & 0.895 & 0.816              & 0.747      & 0.847     & 0.961   \\
High-grade IN    & Seg-Net & 0.894 & 0.812              & 0.757     & 0.881    & 0.913  \\
                 & MedT    & 0.824 & 0.707              & 0.556     & 0.740     & 0.958   \\
\hline
                 & U-Net   & 0.911 & 0.849              & 0.773      & 0.879     & 0.953   \\
Low-grade IN     & Seg-Net & 0.924 & 0.864             & 0.826     & 0.883    & 0.977  \\
                 & MedT    & 0.889 & 0.808             & 0.730     & 0.876     & 0.916   \\
\hline
                 & U-Net   & 0.887 & 0.808              & 0.718      & 0.850     & 0.950   \\
Adenocarcinoma   & Seg-Net & 0.865 & 0.775             & 0.646     & 0.792    & 0.977  \\
                 & MedT    & 0.735 & 0.595             & 0.197     & 0.662     & 0.864   \\
\hline
                 & U-Net   & 0.938 & 0.886              & 0.865      & 0.899     & 0.983   \\
Serrated adenoma & Seg-Net & 0.907 & 0.832              & 0.794     & 0.859    & 0.963  \\
                 & MedT    & 0.670 & 0.509             & -0.043     & 0.896     & 0.544   \\
\hline       
\end{tabular}
\end{table*}

Based on the above results, EBHI-Seg achieved a clear differentiation using deep learning 
image segmentation methods. Image segmentation metrics for different deep learning methods 
are significantly different so that EBHI-Seg can evaluate their segmentation performance.

\section{Discussion}
\label{section:discussion}
\subsection{Discussion of Image Segmentation Results Using Classical Machine Learning Methods}
Six types of tumor differentiation stage data in EBHI-Seg were analyzed using classical 
machine learning methods to obtain the results in Table~\ref{Table:compare}. Base on the 
Dice ratio metrics, $k$-means, MRF and Sobel show no significant differences among the 
three methods around 0.55. In contrast, Watershed metrics are $\sim$0.45 on average, 
which is lower than the above three metrics. OTSU index is around $\sim$0.40 because 
the foreground-background is blurred in some experimental samples and OTSU had a 
difficulty extracting a suitable segmentation threshold, which resulted in undifferentiated 
test results. Precision and Recall evaluation indexes for k-means, MRF, and Sobel are 
also around 0.60, which is higher than those for OTSU and Watershed methods by about 0.20. 
In these three methods, $k$-means and MRF are higher than Sobel in the visual performance 
of the images. Although Sobel is the same as these two methods in terms of metrics, it 
is difficult to distinguish foreground and background images in real images.The segmentation 
results for MRF are obvious but the running time for MRF is too long in comparison with 
other classical learning methods. Since classical machine learning methods have a rigorous 
theoretical foundation and simple ideas, they have been shown to perform well when used for 
specific problems. However, the performance of different methods varied in the present study.

\subsection{Discussion of Image Segmentation Results Using Deep Learning Methods}
In general, deep learning models are considerably superior to classical machine learning 
methods, and even the lowest MedT performance is still higher than the highest accuracy 
of classical machine learning methods. In EBHI-Seg, the Dice ratio evaluation index of 
MedT reaches $\sim$0.75. However, the MedT model size was larger and as a result the 
training time was too long. U-Net and Seg-Net have higher evaluation indexes than MedT, 
both of about 0.88. Among them, Seg-Net has the least training time and the lowest 
training model size. Because the normal category has fewer sample images than other 
categories, the evaluation metrics of the three deep learning methods in this category 
are significantly lower than those in other categories. The evaluation metrics of the 
three segmentation methods are significantly higher in the other categories, with 
Seg-Net averaging above 0.90 and MedT exceeding 0.80.

\section{Conclusion and Future Work}
\label{section:conclusion}

The present stduy introduced a publicly available colorectal pathology image dataset 
containing 5,710 magnified $400\times $ pathology images of six types of tumor 
differentiation stages. EBHI-Seg has high segmentation accuracy as well as good robustness. 
In the classical machine learning approach, segmentation experiments were performed using 
different methods and  evaluation metrics analysis was carried out utilizing segmentation 
results. The highest and lowest Dice ratios are 0.65 and 0.30, respectively. The highest 
Precision and Recall values are 0.70 and 0.90, respectively, while the lowest values are 
0.50 and 0.35, respectively. All three models performed well when using the deep learning 
method, with the highest Dice ratio reaching above 0.95 and both Precision and Recall 
values reaching above 0.90. The segmentation experiments using EBHI-Seg show that this 
dataset effectively perform the segmentation task in each of the segmentation methods. 
Furthermore, there are significant differences among the segmentation evaluation metrics. 
Therefore, EBHI-Seg is practical and effective in performing image segmentation tasks.

 \section*{Data Availability Statement}
 \label{section: data} 
 EBHI-Seg is proved by ``Research Project Ethics Certification'' (No. 202229) from 
 Cancer Hospital of  China Medical University, Shenyang, China. 
 EBHI-Seg is publicly available at: https://doi.org/10.6084/m9.figshare.21540159.v1

\section*{Author Contributions}
 \label{section:contri}
 
 L. Shi: Data preparation, experiment, result analysis, paper writing;
 X. Li: Corresponding author, data collection, medical knowledge;
 W. Hu: Data collection, data preparation, paper writing;
 H. Chen: Data preparation, paper writing;
 J. Chen, Z. Fan, M. Gao, Y. Jing, G. Lu, D. Ma, Z. Ma, Q. Meng, D. Tang: Data preparation; 
 H. Sun: Medical knowledge;
 M. Grzegorzek: Result analysis; 
 S. Qi: Method;
 Y. Teng; Result analysis; 
 C. Li; Corresponding author, data collection, method, experiment, result analysis, paper writing, proofreading.

\section*{Funding}
\label{section:funding}
This work is supported by the ``National Natural Science Foundation of China'' (No. 82220108007) 
and ``Beijing Xisike Clinical Oncology Research Foundation'' (No. Y-tongshu2021/1n-0379).

\section*{Acknowledgements}
\label{section:ack}
We thank Miss. Zixian Li and Mr. Guoxian Li for their important discussion in this work.

 \section*{Conflict of Interest}
 \label{section:dec}
 The authors declare that they have no conflict of interest in this paper.

\bibliography{Shi}

\end{document}